\documentclass[british,english]{article}
\usepackage[T1]{fontenc}
\usepackage[latin9]{inputenc}
\usepackage[letterpaper]{geometry}
\usepackage{amsmath}
\usepackage[authoryear]{natbib}

\makeatletter

\providecommand{\tabularnewline}{\\}

\newcommand{\lyxaddress}[1]{
\par {\raggedright #1
\vspace{1.4em}
\noindent\par}
}
\newenvironment{lyxlist}[1]
{\begin{list}{}
{\settowidth{\labelwidth}{#1}
 \setlength{\leftmargin}{\labelwidth}
 \addtolength{\leftmargin}{\labelsep}
 }}
{\end{list}}

\makeatother

\usepackage{babel}

\begin{document}

\title{Causation, decision theory, and Bell's theorem: a quantum analogue
of the Newcomb problem}
\date{}

\author{Eric G. Cavalcanti}

\maketitle

\lyxaddress{\begin{center}
Centre for Quantum Dynamics, Griffith University, Brisbane, QLD 4111,
Australia
\par\end{center}}
\begin{abstract}
I apply some of the lessons from quantum theory, in particular from
Bell's theorem, to a debate on the foundations of decision theory
and causation. By tracing a formal analogy between the basic assumptions
of Causal Decision Theory (CDT)---which was developed partly in response
to Newcomb's problem--- and those of a Local Hidden Variable (LHV)
theory in the context of quantum mechanics, I show that an agent who
acts according to CDT and gives any nonzero credence to some possible
causal interpretations underlying quantum phenomena should bet against
quantum mechanics in some feasible game scenarios involving entangled
systems, no matter what evidence they acquire. As a consequence, either
the most accepted version of decision theory is wrong, or it provides
a practical distinction, in terms of the prescribed behaviour of rational
agents, between some metaphysical hypotheses regarding the causal
structure underlying quantum mechanics.
\end{abstract}
\tableofcontents{}

\section{Introduction}

Quantum theory has motivated not only radical revisions in our understanding
of physical theory, but in our understanding of other areas of knowledge
which seemed a priori quite dissociated from physics. The most recent
example is the application of the framework of quantum mechanics to
the theory of information processing, leading to the very active fields
of quantum information and computation. One of the precursors to these
more recent developments was the work of John Bell~\citeyearpar{Bell1964},
inaugurating an area of research that has been aptly termed \emph{experimental
metaphysics} by Abner Shimony~\citeyearpar{Shimony1989}. Bell\emph{
}and others since him have shown that a much more intimate relationship
between physics and philosophy is not only possible but fruitful.
After Bell we have come to recognise novel ways in which bare experimental
data can have a direct influence on philosophically-oriented inquiry,
and how such inquiry can indeed be a precursor to new and useful views
on physical theory, which can eventually even lead to new technologies.

Here I argue that the lessons from quantum theory can shed light on
an important debate in the philosophical foundations of decision theory.
This debate started when the philosopher Robert Nozick published \citep{Nozick1969}
a puzzle introduced to him by the physicist William Newcomb---the
so-called Newcomb's paradox or Newcomb's problem. Attempts to solve
this problem and its variants have generated waves of activity in
the philosophy of decision theory. The consensus at present seems
to be around the Causal Decision Theory (CDT) proposed and defended
by Gibbard and Harper~\citeyearpar{Gibbard1978}, Lewis~\citeyearpar{Lewis1981a},
Skyrms~\citeyearpar{Skyrms1982}, among others.

The influence of this problem goes beyond the mere resolution of the
original paradox, and even beyond the foundations of decision theory.
It has also had an effect on debates about the status of causal laws,
since Nancy Cartwright's influential article {}``Causal laws and
effective strategies''~\citep{Cartwright1979}, where she argues
that causal laws cannot be reduced to probabilistic laws of association,
and that they are necessary to distinguish between effective and ineffective
strategies.

For philosophers the importance of this debate is obvious, but should
physicists also care? I believe so. An increasingly popular view among
physicists is the idea that quantum mechanics is a theory of information,
and that quantum states are nothing but concise encapsulations of
subjective probabilities. In this Bayesian view of quantum states
\citep{Caves2002b,Fuchs2003,Fuchs2005}, the gambling commitments
of rational agents should mirror the probabilities assigned by quantum
mechanics to the possible outcomes of possible observations following
a given physical preparation procedure. It is well known that several
problems occur when one tries to attribute underlying causal stories
to quantum phenomena, but there are also many open problems in the
project of quantum Bayesianism. One of them is that, as opposed to
the case of classical probability, there is still no foundationally
attractive way to justify the structure of quantum mechanics from
something as plausible and intuitive as a Dutch-book argument%
\footnote{For a recent account of the progress in that direction, see \citep{Fuchs2009}.%
}. The present work is part of recent attempts to be explicit about
the inclusion of an agent in quantum mechanics---through their \emph{decisions}. 

Decision theory has been used as the basis of a foundational program
started by Deutsch \citeyearpar{Deutsch1999} and further developed
by Wallace \citeyearpar{Wallace2006,Wallace2007} in the context of
the Everett (Many-Worlds) interpretation of quantum mechanics. However,
while that program aims at reconstructing features of quantum mechanics
from decision theory, the present work goes in the opposite direction:
here I will rather argue against a certain theory of decision based
on lessons from quantum mechanics. Furthermore, here we attempt to
go beyond the use of a specific ontological framework, but take as
starting point the epistemological (or more correctly, decision-theoretic)
aspect of the theory. Specific ontological models or frameworks will
be considered, as they will be seen to be fundamental in determining
the decision-theoretic prescriptions according to causal decision
theory. But it is important to emphasise that, in the spirit of experimental
metaphysics, we study the space of possible metaphysical theories
without a particular commitment to one or another beyond what is required
by experimental data. A similar approach can be found in recent studies
of so-called \emph{ontological models} in the context of quantum foundations
\citep{Spekkens2005,Rudolph2006,Harrigan2007b,Harrigan2007c}.

If CDT is right, causal considerations must take priority in generating
the effective probabilities that rational agents should use for their
gambling commitments. According to CDT, rational agents should base
their decision on so-called \emph{causal probabilities}, even when
those are distinct from the subjective (or evidential) probabilities
refined by evidence. This is then not a purely philosophical question
(to use the physicist's jargon for problems which do not have direct
empirical implications), but has a direct consequence in the prescribed
behaviour of rational agents. The debate is not between opposing interpretations
of operationally equivalent theories, but between opposing theories
with conflicting prescriptions. If quantum mechanics is nothing but
a theory of information, and if probabilities are nothing but the
gambling commitments of rational agents, and if one of the lessons
of quantum mechanics is the demise of causal concepts in the prescription
of those probabilities, it should be a problem for the Bayesian-inclined
physicist that the most popular version of decision theory attempts
to put causation at the root of an agent's effective probability assignments.
This work can thus be seen as an attempt to make decision theory \textquotedbl{}safe\textquotedbl{}
for quantum Bayesianism%
\footnote{Although I should mention that I disagree with aspects of the quantum
Bayesianism defended by Fuchs \emph{et al}. In particular, I think
this approach would be much more productive if it focused on the program
of attempting to derive as much of quantum mechanics as possible from
information- or decision-theoretic principles, and abstained from
opposing specific ontological models. After all, everyone can agree
that there are such things as subjective probabilities associated
to experimental outcomes (even if some may disagree on whether or
not there are \emph{also }other kinds of probabilities), and thus
everyone could find it interesting to know whether the probabilities
prescribed by quantum mechanics can be derived from information-theoretic
principles (even if they could \emph{also} be derived from a constructive
ontological model).%
}.

The main goal of this paper is to make this concern explicit and trace
a formal parallel between CDT and the class of Local Hidden Variable
(LHV) theories which are discussed in the context of Bell's theorem~\citep{Bell1964,Bell1987}.
This parallel seems to lend support to Bayesian Decision Theory (BDT)
over CDT. I will show that, in general, CDT represents a limitation
on the space of effective probability assignments available to an
agent---just as is the case with LHV\textquoteright{}s in quantum
mechanics---and this can under certain conditions render an agent
incapable of adjusting their effective probabilities (even if they
adjust their subjective probabilities) to match arbitrarily well to
some possible observations, no matter what evidence they accumulate.
In fact, I will argue that even in some routine quantum experiments,
a causal decision theorist would be forced, under a plausible analysis
of the prescriptions of CDT, to bet against some observed predictions
of quantum mechanics. As a result, either CDT is wrong, or it surprisingly
provides a practical distinction, in terms of the prescribed behaviour
or rational agents, between some causal hypotheses underlying quantum
mechanics.

\section{Newcomb's problem}

The original Newcomb problem \citep{Nozick1969} is as follows. You
are in a room with two boxes, labelled $1$ and $2$. Box $2$, you
can see, contains a thousand dollars. Box $1$ is closed. A Predictor,
in whom you have high confidence to be able to predict your own choices
(she has accurately predicted your choices in several similar situations
in the past, say), proposes the following game to you: you can either
choose to take both boxes in front of you, or choose only Box $1$.
She tells you that before you entered the room, she predicted what
you would do. She also tells you that if she predicted that you were
going to take only Box $1$, she has put a million dollars inside.
If she predicted you would take both boxes, she has put nothing in
it. What should you do?

Bayesian decision theory prescribes the maximisation of expected utility.
In a general decision situation, we denote by $A_{i},\, i\in\{1,2,...,n\}$
the several actions available to the agent, and by $O_{j},\, j\in\{1,2,...,m\}$
the possible outcomes. The agent ascribes to each pair action/outcome
a numerical utility $u(A_{i},O_{j})$ and a conditional probability
$P(O_{j}|A_{i})$. Since by assumption the $O_{j}$ form a complete
set of mutually exclusive events, $\sum_{j}P(O_{j}|A_{i})=1.$ The
Bayesian (or evidential) expected utility of each action is therefore
\begin{equation}
EU(A_{i})=\sum_{j}P(O_{j}|A_{i})\, u(A_{i},O_{j}).\label{eq:EU_E(A)}\end{equation}

Now denote by $O_{1}$ the event that you find a million dollars in
Box $1$, and by $A_{1}$ and $A_{2}$ your action of choosing Box
$1$ and both boxes, respectively. Your information about the Predictor's
efficacy is represented by the fact that your conditional probabilities
are such that $P(O_{1}|A_{1})\gg P(O_{1}|A_{2})$. The Bayesian expected
utilities are \begin{eqnarray}
EU(A_{1}) & = & P(O_{1}|A_{1})\,1,000,000+[1-P(O_{1}|A_{1})]\,0\nonumber \\
EU(A_{2}) & = & P(O_{1}|A_{2})\,1,001,000+[1-P(O_{1}|A_{2})]\,1,000.\label{eq:Newcomb_1}\end{eqnarray}
We assume that the utilities of each pair action/outcome are just
the money values received by the agent. This is not a restrictive
assumption, as one could adapt the money values if necessary such
that the utilities are as required. It is easy to see that if $P(O_{1}|A_{1})$
is sufficiently larger than $P(O_{1}|A_{2})$, the expected utility
of choosing one box will be larger than the expected utility of choosing
both. In the circumstance of Newcomb's problem, therefore, EDT advises
you to choose one box.

\section{Causal Decision Theory\label{sec:Causal-Decision-Theory}}

Although the above answer seems correct at first, a second argument
soon comes to mind: whatever I do, the first box either contains a
million dollars or it doesn't. This fact was settled in the past and
is beyond the causal influence of my present choice. Nothing that
I can do now will change the contents of Box $1$. But regardless
of whether it contains a million dollars or nothing, I'll be better
off taking the extra thousand. Therefore I should take both boxes.
This is called the \emph{dominance argument,} since one choice seems
to dominate the other no matter what outcome\emph{ }obtains\emph{.}

Causal Decision Theory was developed as an attempt to formalise the
intuition behind the dominance argument. Gibbard and Harper \citeyearpar{Gibbard1978}
claimed that the expected utility of an action should be calculated
from the probabilities of counterfactuals, as opposed to the conditional
probabilities that figure in \eqref{eq:EU_E(A)}. Under the evaluation
of the probabilities of counterfactuals favoured by Gibbard and Harper,
this principle of utility maximisation prescribes the desired two-boxing
strategy in Newcomb's problem. However, there are different possible
interpretations for counterfactuals, and this strategy is therefore
ambiguous. Horgan \citeyearpar{Horgan1981}, for example, argues that
a \textquotedbl{}backtracking\textquotedbl{} analysis of the counterfactuals
leads to the prescription of one-boxing.

There is a consensus, however, on the final mathematical form of the
utility formula defended by CDT \citep{Lewis1981a,Skyrms1982,Armendt1986}.
According to these authors, the correct quantity to be maximised in
a decision situation is the causal expected utility\begin{equation}
CEU(A_{i})=\sum_{j}\left[\sum_{\lambda}P(K_{\lambda})\, P(O_{j}|A_{i};K_{\lambda})\right]u(A_{i},O_{j}),\label{eq:EU_C(A)}\end{equation}
where the $K_{\lambda}'s$ represent the \textquotedbl{}dependency
hypotheses\textquotedbl{} \citep{Lewis1981a}\textbf{ }available to
the agent, and I use the notation of separating the contemplated actions
$A_{i}$ from other propositions with a semicolon. A dependency hypothesis,
according to Lewis \citeyearpar{Lewis1981a}, is \textquotedbl{}a
maximally specific proposition about how the things {[}the agent{]}
cares about do and do not depend causally on his present actions\textquotedbl{}.
On Skyrms' \citeyearpar{Skyrms1982} account, the propositions $K_{\lambda}$
represent the possible \textquotedbl{}causal propensities\textquotedbl{}
that are objectively instantiated in the world. Lewis \citeyearpar{Lewis1981a}
reads Skyrms as describing them as {}``maximally specific specifications
of the factors outside the agent's influence (at the time of decision)
which are causally relevant to the outcome of the agent's action''.

The important thing to note, regardless of the interpretational fine
print, is that formally the expression in brackets represents a {}``causal
probability'' defined as \begin{equation}
P_{c}(O_{j}|A_{i})\equiv\sum_{\lambda}P(K_{\lambda})\, P(O_{j}|A_{i};K_{\lambda})\label{eq:causal probability}\end{equation}
which is in general distinct from the conditional probability, which
can be decomposed as (assuming that there exists a joint probability
distribution for the $K_{\lambda}$ and $O_{j}$), \begin{equation}
P(O_{j}|A_{i})=\sum_{\lambda}P(K_{\lambda},O_{j}|A_{i})=\sum_{\lambda}P(K_{\lambda}|A_{i})\, P(O_{j}|A_{i};K_{\lambda}),\label{eq:evidential probability}\end{equation}
 The distinction, as is clear in this notation, is that in evaluating
the causal probabilities, one ignores any statistical correlation
between the dependency hypotheses and the actions $A_{i}$. In general,
however, such correlations may exist in the expression \eqref{eq:evidential probability}
for the conditional probability. I will call a \emph{Newcomb-type
problem} any decision problem for which the prescriptions of CDT and
BDT disagree. A Newcomb-type problem can be concocted, by appropriate
choice of utilities, in any circumstance where causal probabilities
differ from the conditional probabilities.

For those unfamiliar with this formalism, and who lack a clear intuition
for two-boxing in the original Newcomb problem, this may sound like
an unjustifiable move. So as not to be unfair to the causal decision
theorist, I will present the kind of case where the intuitions favour
CDT the most: the {}``medical Newcomb problems''. 

In a common version of a medical Newcomb problem%
\footnote{Here I have adapted an example from Price \citeyearpar{Price1986}.%
}, it is found that smoking does not cause lung cancer. Instead, it
is discovered that the correlation between smoking and lung cancer
is a spurious one, arising from the existence of a common cause, a
certain gene $G$. The presence of this gene is correlated with smoking,
so that it occurs in, say, 20\% of smokers but only in 2\% of nonsmokers.
It is also highly correlated with lung cancer: almost all bearers
of this gene develop lung cancer if they don't die earlier of other
causes, and the likelihood of a non-bearer to develop lung cancer
is negligible. Given the presence (or absence) of the gene, however,
smoking is rendered uncorrelated with lung cancer.

Now imagine that Fred knows all this and is trying to decide whether
or not to smoke (or continue smoking). He likes smoking, but the prospect
of cancer outweighs his desire for smoking. Suppose his desires, and
the (evidential) conditional probabilities he takes the available
evidence to imply in his case, are as represented on Table \ref{tab:The-smoking-gene}.

\begin{table}
\centering{}\begin{tabular}{|c|c|c|}
\hline 
 & Cancer ($C$) & No cancer ($\neg C$)\tabularnewline
\hline
\hline 
Smoking ($S$) & (0.2, -99) & (0.8, 1)\tabularnewline
\hline 
Not Smoking ($\neg S$) & (0.02, -100) & (0.98, 0)\tabularnewline
\hline
\end{tabular}\caption{{\small The smoking gene scenario. The ordered pairs $(p,u)$ represent
the conditional probabilities $p=P(O_{j}|A_{i})$ and pay-offs $u=u(A_{i},O_{j})$
for each choice $A_{i}$ and outcome $O_{j}$.\label{tab:The-smoking-gene}}}

\end{table}

Given these data, the evidential expected utility of smoking is $EU(S)=-19$
and that of not smoking is $EU(\neg S)=-2.$ BDT therefore advises
Fred not to smoke. Within causal decision theory, on the other hand,
and taking the presence of the gene as the dependency hypothesis in
\eqref{eq:EU_C(A)}, $P(C|S;G)=P(C|\neg S;G)$ and $P(C|S;\neg G)=P(C|\neg S;\neg G),$
therefore whatever Fred's prior beliefs $P(G)$ about his genetic
endowment are, $CEU(S)>CEU(\neg S),$ and CDT advises him to smoke,
as is intuitively the correct prescription for most people.

Although this example seems to strongly support CDT, there are defences
available which allow BDT to achieve the same prescription. I will
return to these in Section \ref{sec:Communicated-vs.-non-communicated}.

\subsection{Regions of causal influence\label{sub:Regions-of-causal}}

There is an important point to emphasise. Causal decision theory assumes
not only that there exists a distribution over a complete specification
of the causal propensities in general but also, although perhaps less
explicitly, that these dependency hypotheses screen off the correlations
between an action and all events which are outside its causal influence.
Formally, this means that \begin{equation}
P(O_{j}|A_{i};K_{\lambda})=P(O_{j}|A_{i'};K_{\lambda})=P(O_{j}|K_{\lambda})\label{eq:CDT_"locality"}\end{equation}
for all $(i,\, i')$, whenever $O_{j}$ is outside the causal influence
of $A_{i}.$ Therefore the causal probabilities given by Eq. \eqref{eq:causal probability}
reduce in those cases to \begin{equation}
P_{c}(O_{j}|A_{i})\equiv\sum_{\lambda}P(K_{\lambda})\, P(O_{j}|K_{\lambda}).\label{eq:Newcomb_P_c}\end{equation}
The justification for this is the belief that, for example, if I knew
about the gene, my chance of cancer would not be statistically correlated
with my choice of smoking; if I knew what the prediction was, the
money in the box would not be correlated with my choice of picking
one or both. If this assumption isn't made, nothing prevents a direct
dependence between those outcomes and my choices. In other words,
if it is to serve the purpose it is meant to---i.e., to prescribe
\textquotedbl{}two-boxing\textquotedbl{} in Newcomb-type problems---CDT
necessarily needs an account of what are the sets of events {}``inside''
and {}``outside'' the causal influences of an action. In Newcomb
problems, these are typically events that happened in the past of
the action. In general, however, taking into account relativity, this
region could be expanded to include all events outside the future
light cone of the action---both the past light cone and space-like
separated regions. In any case, the causalists need an account of
what these sets are supposed to be which can be used consistently
throughout all decision problems.

Of course, in particular problems the reasoning behind the assignment
of event $O$ as outside the causal influence of action $A$ may not
be due to a fundamental physical constraint such as the speed of light,
but due to other constraints that arise out of an understanding of
the physical situation of the problem. For example, relativity does
not prohibit that a choice we make could change our genes, but this
possibility is disregarded due to our understanding of genetics. Therefore
the term \textquotedbl{}regions of causal influence\textquotedbl{}
may not necessarily refer to actual space-time regions, but to more
general sets of events. In any case, it seems that the regions of
causal influence should be taken from our best scientific theories
about the physical situation underlying a decision scenario.

This formalism seems to miss an important issue, however. What if
some of the outcomes in a decision situation are outside the agent's
influence, but some are not? Let us suppose we have a set of outcomes
$a_{j}$ which the agent takes to be within the causal influence of
the choices $A_{i},$ and another set of outcomes $b_{l}$ which are
taken to be outside the agent's possible causal influence. Now suppose
the pay-offs of a decision situation depend on both of these events.
CDT then needs a joint causal probability $P_{c}(a_{j},b_{l}|A_{i}).$
The fact that we are now considering two variables isn't important---this
should still be given by the obvious generalisation of \eqref{eq:causal probability},\begin{equation}
P_{c}(a_{j},b_{l}|A_{i})\equiv\sum_{\lambda}P(K_{\lambda})\, P(a_{j},b_{l}|A_{i};K_{\lambda}).\label{eq:causal_p_OW_1}\end{equation}
We can always decompose the conditional probability inside the summation
as $P(a_{j},b_{l}|A_{i};K_{\lambda})=P(a_{j}|A_{i};K_{\lambda})P(b_{l}|A_{i};a_{j},K_{\lambda})$.
And, since each $K_{\lambda}$ represents \textquotedbl{}a maximally
specific proposition about how the things {[}the agent{]} cares about
do and do not depend causally on {[}the agent's{]} present actions\textquotedbl{},
this can be simplified to $P(a_{j},b_{l}|A_{i};K_{\lambda})=P(a_{j}|A_{i};K_{\lambda})P(b_{l}|K_{\lambda}).$
The reason, as before, is that by assumption the $b_{l}'s$ are not
causally dependent on the $A_{i}'s,$ only statistically correlated
via some common cause, maximally specified by $K_{\lambda}.$ And
since the $a_{j}'s$ are causally dependent on the $A_{i}'s$, the
$b_{l}'s$ cannot depend directly on the $a_{j}'s$ either. Otherwise
by influencing $a_{j}$ the agent could influence $b_{l},$ contrary
to the assumption. Substituting this expression on \eqref{eq:causal_p_OW_1},
the causal probabilities in this scenario become\begin{equation}
P_{c}(a_{j},b_{l}|A_{i})=\sum_{\lambda}P(K_{\lambda})\, P(a_{j}|A_{i};K_{\lambda})P(b_{l}|K_{\lambda}).\label{eq:causal_prob_OW_final}\end{equation}
Given the utilities $u(A_{i},a_{j},b_{l}),$ the causal expected utility
which generalises Eq. \eqref{eq:EU_C(A)} is\begin{equation}
EU(A_{i})=\sum_{j,l}\left[\sum_{\lambda}P(K_{\lambda})\, P(a_{j}|A_{i};K_{\lambda})P(b_{l}|K_{\lambda})\right]u(A_{i},a_{j},b_{l}).\label{eq:general_EU_C}\end{equation}
We could also consider a situation involving not only one but two
agents. We could interpret $b_{l}$ as being the outcomes observed
by a second agent, who has at their disposal a number of possible
choices $B_{k}.$ Of course, the $b_{l}$ are within the causal influence
of $B_{k},$ and therefore the more general causal probabilities are
given by $P_{c}(a_{j},b_{l}|A_{i},B_{k})=\sum_{\lambda}P(K_{\lambda})\, P(a_{j}|A_{i},B_{k};K_{\lambda})P(b_{l}|B_{k};K_{\lambda})$
if the actions $B_{k}$ can directly causally influence the outcomes
$a_{j}$ (by being in their past, say), or \begin{equation}
P_{c}(a_{j},b_{l}|A_{i},B_{k})=\sum_{\lambda}P(K_{\lambda})\, P(a_{j}|A_{i};K_{\lambda})P(b_{l}|B_{k};K_{\lambda})\label{eq:prob_causal_A_B}\end{equation}
if they cannot.

\subsection{Evidential and effective probabilities\label{sub:Evidential-and-effective}}

I should stress the fact that for the causal decision theorist there
are two kinds of probabilities: those that represent their evidence,
their degrees of belief, about the possible outcomes conditional on
the performance of each action available to them, and the probabilities
that they should use to ground their decisions. I will call the former
kind of probability the {}``evidential'' or {}``subjective'' probabilities,
and the latter kind the agent's {}``effective'' probabilities. For
evidential decision theorists, these two probabilities coincide: they
ground their actions on their subjective conditional probabilities.
For causal decision theorists they come apart in scenarios such as
Newcomb's. They can also come apart, I will argue in the next section,
in actually feasible scenarios involving quantum experiments.

It is important however to remind the reader that the causal decision
theorist does not deny the existence or meaning of the subjective
probabilities. They indeed believe in the same subjective conditional
probabilities. They believe that the one-boxers in Newcomb's original
problem are more likely to come out richer than the two-boxers, but
they believe they come out richer for the wrong reasons. As Lewis
\citeyearpar{Lewis1981b} puts it:
\begin{quote}
{\small They have their millions and we have our thousands, and they
think this goes to show the error of our ways. They think we are not
rich because we have irrationally chosen not to have our millions.
We reply that we never were given any choice about whether to have
a million... The reason why we are not rich is that the riches were
reserved for the irrational.}{\small \par}
\end{quote}
This implies that while the causalist believes in the same conditional
probabilities as the evidentialist, with apparently the same interpretation,
the causalist also believes that those should not ground their decisions.
Instead, they take their effective probabilities to be the causal
probabilities \eqref{eq:causal probability} (which implicitly mean
in general, as I argued, \eqref{eq:causal_prob_OW_final}).

It is important to note that the effective probabilities of a causal
decision theorist need not be updated by evidence in the same manner
as the evidential probabilities. For example, in the original Newcomb
scenario, repeatedly playing the game and observing strong correlations
between one-boxing and a million dollars, and two-boxing and a thousand
dollars, could influence the agent's evidential probabilities, which
should be properly updated through Bayes' rule. But it could not change
the agent's causal probabilities; by assumption the contents of the
box are outside the agent's causal influence, and the correlations
are (by assumption) explained in that case by the existence of a common
cause for the agent's choices as well as the contents of the box.
This refusal to change his decisions even in the face of the winnings
of the one-boxer is what is illustrated by the Lewis quotation above.

To be sure, it is important that the agent's causal story can \emph{explain
}the evidential correlations. For example, in the smoking gene scenario,
the assumption that the gene causes both smoking and lung cancer explains
the correlation between the two. In Newcomb's problem the Predictor's
ability explains the correlations between the agent's choices and
the contents of the box. If some evidence arises that is incompatible
with the causal story held by the agent, then of course the agent
would be compelled to revise their causal hypotheses. In general,
however, if the agent's causal hypotheses provide a causal explanation
for the observed correlations, then the correlations cannot suggest
a change in those hypotheses. Correlations which are already expected
or predicted by the causal hypotheses cannot present any new information
to modify those.

If the causal probabilities were updated in the same manner as the
subjective probabilities, then CDT and BDT would tend to agree in
the long run, and the conflict between the two would disappear. The
idea that causal probabilities should tend to agree with subjective
probabilities (under some interpretation of causation) is a position
that can and has been defended, for example, by Price \citeyearpar{Price1991}.
The arguments in this paper will be directed towards those who are
not swayed by Price's program and maintain that there can be differences
between causal and subjective or evidential probabilities (and thus
between the prescriptions of CDT and BDT).

One of the reasons Newcomb's problem is so controversial, I believe,
is that Newcomb-type problems have been generally purely hypothetical
and quite far-fetched scenarios. The adherence to each contender theory
never had to be tested in practical decision situations. In the following
I will make a parallel that should serve to provide a feasible example.

\section{The parallel with Bell's theorem}

The main argument of this paper is based on a formal and conceptual
analogy between the causal probabilities as applied to a Newcomb scenario
and the probabilities prescribed by a local hidden variable theory
in the context of quantum mechanics. In this section I will present
this analogy. 

Let me first introduce the relevant notation. Alice and Bob are agents
who have at their disposal a number of possible measurements ($A_{i}$
for Alice, $B_{k}$ for Bob) each of which with a number of possible
outcomes $a_{j}^{i},$ $b_{l}^{k},$ respectively (we could introduce
other agents if necessary, of course, but two will be sufficient for
our purposes). We will henceforth constrain ourselves for simplicity
to the cases in which all measurements have the same number of outcomes,
and identify $a_{j}^{i}=a_{j}^{i'}=a_{j}$ for all $i,\, i',$ and
similarly for the $b_{l}^{k}.$ Alice assigns a subjective conditional
probability $P(a_{j}|A_{i})$ to each possible outcome of each possible
experiment.

A general hidden variable theory for the phenomena observed by Alice
and Bob consists of a probability distribution over the elements $\lambda$
of a set of hidden variables $\Lambda,$ together with a distribution
for the possible experimental outcomes (given $\lambda$) which reproduces
the observed statistics, i.e.\begin{equation}
P_{HV}(a_{j},b_{l}|A_{i},B_{k})\equiv\sum_{\lambda}P(\lambda)P(a_{j},b_{l}|A_{i},B_{k};\lambda).\label{eq:HV_general}\end{equation}

These variables are supposed to represent a sufficiently complete
specification of physical variables that are causally relevant to
the outcomes of the experiments under study. The requirement that
they be causally relevant is translated within classical relativistic
mechanics to the requirement that they must be specified in the past
light cones of those experiments. The important point is that they
must be necessarily specified in some region of space-time which can
{}``causally influence'' the experiments, according to some theory
of causation. There is at the outset an important assumption used
in the equation above: 
\begin{lyxlist}{00.00.0000}
\item [{\textbf{Statistical~independence.}}] The hidden variables are
statistically independent of the choice of experiments made by Alice
and Bob, i.e., $P(\lambda|A_{i},B_{k})=P(\lambda)$.
\end{lyxlist}
Some authors call this the {}``free will'' assumption, or perhaps
\textquotedbl{}no-retrocausality\textquotedbl{} assumption. I prefer
not to use the term \textquotedbl{}free will\textquotedbl{} so as
not to presuppose an interpretation of the concept of free will which
precludes an account in which it is compatible with determinism. And
I don't favour the term \textquotedbl{}no-retrocausality\textquotedbl{}
because although a dependence of the hidden variables on those experimental
settings would be essentially indistinguishable from backwards causation
from the agent's perspective, it is logically possible for statistical
independence to fail even when there is no actual backwards causation.

This assumption seems to be justified by the fact that these choices
are completely arbitrary. They could be made as a function of the
intensity of a measurement of the cosmic background radiation, or
at the whim of the {}``free-willed'' experimentalists. The variety
and arbitrariness of possible sources seem to imply that they cannot
be correlated with the variables which are causally relevant to this
particular laboratory experiment. 

The extra assumption that will lead to a \emph{local} hidden variable
theory is that there exists some such sufficient specification of
variables that renders the probability of an event $E_{1}$ uncorrelated
with that of an event $E_{2}$, when the event $E_{2}$ is outside
the region of causal influence of $E_{1}.$ In a typical Bell scenario,
this is usually translated as the requirement that Alice's and Bob's
experiments are in space-like separated regions so that the following
holds:
\begin{lyxlist}{00.00.0000}
\item [{\textbf{Local\ causality.}}] $P(a_{j}|A_{i},B_{k};b_{l},\lambda)=P(a_{j}|A_{i};\lambda)$,
and similarly for Bob.
\end{lyxlist}
With this assumption we obtain what is called a local hidden variable
(LHV) model for this experimental scenario: \begin{equation}
P_{LHV}(a_{j},b_{l}|A_{i},B_{k})\equiv\sum_{\lambda}P(\lambda)P(a_{j}|A_{i};\lambda)P(b_{l}|B_{k};\lambda).\label{eq:LHV}\end{equation}

The analogy should start to be clear. Eqs. \eqref{eq:prob_causal_A_B}
and \eqref{eq:LHV} are formally identical. I will now analyse in
some more detail the Newcomb scenario discussed above to make sure
that the conceptual analogies are also clear. The scenario involves
not only the choices and direct observations of the agent (let us
say Alice is this agent) but also the actions and the outcomes of
the actions of the Predictor, which happen in a region outside the
causal influence of Alice. In the Newcomb scenario, that is translated
as the fact that the Predictor's actions are in the past of Alice's
choice. Even if there is no actual Predictor (as in the smoking gene
scenario) we can always model the situation by imagining those events
outside Alice's causal influence as being the actions performed and
outcomes observed by an agent, with trivial actions/outcomes where
necessary. This will allow a more explicit and direct comparison between
the two models, without modifying in any way the prescriptions of
causal decision theory.

The assumption of \textquotedbl{}statistical independence\textquotedbl{}
in a general hidden variable model, which leads to Eq. \eqref{eq:HV_general},
is formally and conceptually equivalent to the assumption that the
effective causal probabilities are given by an average over the unconditional
probabilities of the causal propensities $K_{\lambda},$ Eq. \eqref{eq:causal probability}
(and \eqref{eq:causal_p_OW_1}). The assumption of \textquotedbl{}local
causality\textquotedbl{} which leads to a LHV model is formally and
conceptually equivalent to the assumption that outcomes outside Alice's
regions of causal influence are screened by the causal propensities
$K_{\lambda},$ Eq. \eqref{eq:CDT_"locality"}. 

Thus in the Newcomb problem we can model the situation by imagining
that Bob is the Predictor. He chooses a trivial available action ($B_{1}$)
to put a million dollars inside the closed box if and only if he predicts
Alice will pick only the closed box. He will base his prediction on
his knowledge of some variables $\lambda,$ necessarily specified
in his own past light cone. These variables, he believes, will be
correlated with Alice's future choice. He will plug these variables
into an algorithm, say, and observe outcomes corresponding to the
prediction that Alice will ($b_{1}$) pick the closed box only or
($b_{2}$) pick both boxes. He tells this whole story to Alice, as
usual, and asks her to make her decision. She can either choose to
($A_{1}$) pick the closed box or ($A_{2}$) pick both boxes. The
outcomes associated to her choice, however, cannot be whether or not
the box contains a million dollars, since that is not under her direct
causal influence. Those outcomes are (were) under Bob's causal influence,
not hers. This is the reasoning that leads to Eq. \eqref{eq:Newcomb_P_c}
and which allows CDT to prescribe two-boxing. So I will use the trivial
outcomes: ($a_{1}$) she opens only one box or ($a_{2}$) she opens
both boxes. Let us stipulate that the pay-offs depend on the explicit
actions (the $a_{j}$'s), not on the choices, so that the utilities
attributed to each possible pair of outcomes are as given by Table
\ref{tab:Pay-off-matrix}.

\begin{table}
\begin{centering}
\begin{tabular}{|c|c|c|}
\hline 
 & $b_{1}$= \textquotedbl{}Bob predicts $a_{1}$\textquotedbl{} & $b_{2}$= \textquotedbl{}Bob predicted $a_{2}$\textquotedbl{}\tabularnewline
\hline
\hline 
$a_{1}$= \textquotedbl{}Alice takes Box 1\textquotedbl{} & 1,000,000 & 0\tabularnewline
\hline 
$a_{2}$= \textquotedbl{}Alice takes both\textquotedbl{} & 1,001,000 & 1,000\tabularnewline
\hline
\end{tabular}
\par\end{centering}

{\footnotesize \caption{{\small \label{tab:Pay-off-matrix}Pay-off matrix in a Newcomb-type
problem}}
}
\end{table}

What are the causal probabilities that Alice should use in her decision?
The scenario described above is precisely the type that led to Eq.
\eqref{eq:prob_causal_A_B}, identifying $K_{\lambda}\leftrightarrow\lambda:$
\begin{equation}
P_{c}(a_{j},b_{l}|A_{i},B_{k})=\sum_{\lambda}P(\lambda)P(a_{j}|A_{i},\lambda)P(b_{l}|B_{k},\lambda).\label{eq:causal_prob_Bell}\end{equation}
Since there is only a trivial choice for Bob, and using the usual
assumption that Alice is always able to carry out her chosen strategy,
$P(a_{j}|A_{i},\lambda)=\delta_{ij},$ where the Kronecker delta is
defined as $\delta_{ij}=1$ if $i=j$ and $0$ if $i\neq j.$ This
simplifies the equation above to $P_{c}(b_{l}|A_{i})=\sum_{\lambda}P(\lambda)P(b_{l}|\lambda),$
identical to Eq. \eqref{eq:Newcomb_P_c}, and which leads to the prescription
of two-boxing as already argued.

\subsection{Consequences of the analogy\label{sub:Consequences-of-the}}

This analogy can lead to two lines of criticism of CDT, both based
on the fact that the causal probabilities can disagree with the quantum
probabilities. The main line of criticism being pursued in this paper%
\footnote{The other will be mentioned in a footnote in Section \ref{sub:The-Bell-game}.%
} is to show that proper consideration of the various alternative causal
hypotheses proposed as explanations for the quantum correlations will
lead the causalist to bet against the outcomes predicted by quantum
mechanics for certain feasible experimental situations. This will
make use of a game that can be actually set up in a standard quantum
optics laboratory.

I will argue that given the analogy between causal probabilities and
local hidden variable models, the causalist should consistently give
some weight to a local hidden variable model as their effective probabilities
in the quantum-mechanical experiments. Since real experiments routinely
violate these assumptions, through violation of Bell-type inequalities
\citep{Bell1964,Bell1987}, the causalist would stand to lose money.
I will first introduce a decision scenario with no mention of quantum
mechanics, so as to make the discussion of the causalist's decision
simpler to follow. Later I will show how this scenario can be set
up with a pair of entangled quantum systems, and the reasoning that
leads the causalist to bet against quantum mechanics in that game.

\subsubsection{The marble boxes game}

Alice enters a room which has a collection of $N$ black boxes in
a long row. They are all labelled sequentially by integers from $1$
to $N$. These boxes have two buttons each: a red button and a green
button. The boxes are closed and completely sealed from external influences,
as far as she can tell. Each box behaves as follows: when Alice first
presses one of the buttons, a marble of the same colour as the button
she pressed is released from a small circular opening. Each marble
has a symbol, which is either `$+1$' or `$-1$'. Once a marble has
emerged from a box, further button presses on that box do nothing. 

She then hears a familiar voice coming out of a monitor in one corner
of the room. The face is also familiar: it is Bob, the famous game-show
host, who proposes a game to Alice. Bob says he has a set of boxes
that work by the same mechanism in his studio in Brisbane, half a
world away from Alice's location in Amsterdam. His boxes are also
labelled sequentially from 1 to $N$, so that each box at Alice's
location has a corresponding one at Bob's. He tells Alice that she
can choose to press either a red button or a green button at her will
on each of her remaining boxes. He tells her that these boxes can
\emph{predict} whether she will press the red or green buttons. But
it's not so simple, he admits. The prediction isn't perfect, but it's
better than even odds. Moreover, it's encoded in a certain way so
that the prediction of each individual choice cannot really be retrieved,
but only inferred from the correlations between the balls coming out
of her boxes and those of his own boxes.

Alice's friend Charlie is in Bob's studio to guarantee Alice can trust
the game. Bob explains that Charlie will press a red or green button
at random in the remaining of his boxes, simultaneously with Alice,
and similar marbles will emerge from Bob's boxes.

For each pair of marbles from correspondingly-numbered boxes, he will
multiply the numbers printed on them (to give a product of either
+1 or -1), and keep a record of the pair of colours with a code \textquotedbl{}$a_{r}b_{r}$\textquotedbl{}
for red-red, \textquotedbl{}$a_{r}b_{g}$\textquotedbl{} for red-green,
etc. After all the buttons are pressed and all the marbles are released,
he will finally take the averages $\langle a_{r}b_{r}\rangle,$ etc,
of each of these values over the collection of boxes (where $\langle a_{r}b_{r}\rangle$
denotes the average of the $+1$ and $-1$ values of all pairs that
came up red-red, and so on) and plug them into the formula \begin{equation}
\langle F\rangle=\langle a_{r}b_{r}\rangle+\langle a_{r}b_{g}\rangle+\langle a_{g}b_{r}\rangle-\langle a_{g}b_{g}\rangle.\label{eq:Bobs_formula}\end{equation}
If one or more pairs of button colours ($\langle a_{r}b_{r}\rangle$,
$\langle a_{r}b_{g}\rangle$, etc) does not occur, the corresponding
value will be zero. Bob says that if she chooses to play, and the
value of $\langle F\rangle$ is larger than $2.8,$ she wins a million
dollars. If $\langle F\rangle$ is less than or equal to 2.8 she goes
home empty-handed. Alternatively, she can choose not to play and take
home a thousand dollars, risk-free.

\subsubsection{Mechanism underlying the marble boxes game\label{sub:Mechanism-underlying-the}}

Bob explains the reasoning behind the claim that the boxes can predict
Alice's actions. Inside each box, there's already a pair of balls---one
red, one green---with numbers written on them. Let's consider a single
pair of boxes, and call the numbers on the two balls in each box $a_{r}$,
$a_{g}$, and $b_{r}$, $b_{g}$. Recall that each of these numbers
can only be $+1$ or $-1$, and consider the sum $b_{r}+b_{g}$ and
difference $b_{r}-b_{g}$ of the numbers on Bob's marbles. If both
marbles have the same value, then their difference is $0$ and their
sum is either $+2$ or $-2$; if the marbles have opposite values,
their sum is $0$ and their difference is $+2$ or $-2$. Now consider
the formula $F=a_{r}(b_{r}+b_{g})+a_{g}(b_{r}-b_{g}).$ If Bob's marbles
have the same value, we have $F=\pm2a_{r}$; if they have opposite
values we have $F=\pm2a_{g}.$ Thus, in either case, $F$ can only
take on the value $+2$ or $-2.$ 

Multiplying out the above formula $F=a_{r}b_{r}+a_{r}b_{g}+a_{g}b_{r}-a_{g}b_{g}=\pm2$
we see how the above analysis for a single box puts constraints on
the value for $\langle F\rangle$ we should expect from the set of
all boxes. That is, if Alice's and Charlie's choices were really random,
or at least not correlated with the mechanism behind the boxes, then
the expectation value of $F$ for the group of boxes would also be
at most $2.$ After all, each average in that sum would be taken over
the same ensemble of marbles. But it turns out that this average value
is, in practice, always very close to the magical number of $2\sqrt{2}\approx2.828.$
The explanation for that weird situation, Bob guarantees, is that
the boxes are created from a single source, and at that time the internal
mechanism of the boxes somehow \textquotedbl{}knows\textquotedbl{}
what buttons are going to be pressed, and prints the numbers on each
pair of balls so as to ensure that $\langle F\rangle\approx2\sqrt{2}$.
In other words, the mechanism prepares a different distribution of
numbers on the pair of marbles for each pair of buttons to be pressed.
Therefore each of the expectation values in Eq. \eqref{eq:Bobs_formula}
can assume independent values, and the reasoning that restricts the
value of $F$ doesn't apply. In fact, without the assumption that
the distributions for each pair of buttons are the same, the value
of $\langle F\rangle$ could logically be anything between $-4$ and
$+4$, since each term in that sum could be anything between $-1$
and $+1$. 

Alice isn't convinced. ``What if we look inside? Then Charlie and
I would be able to find out what the outcomes are supposed to be,
and we could obtain information about what the boxes have predicted.
Then we could do otherwise. So how's that possible?'' Bob replies
that if they open the boxes, the marbles are destroyed instantaneously.
And when they press a button to release one of the balls, the other
ball is similarly destroyed. So they can never really find out what
the prediction was; they can only recognise by the above reasoning
that the boxes somehow knew what they were going to choose. This,
Bob explains, is the only really secure way to avoid the information
about the prediction reaching Alice, and her doing something to prevent
it from happening. Otherwise, as Alice correctly pointed out, it would
be impossible for these boxes to do what they do.

``Now even if you don't believe that the numbers on the marbles are
already there'', Bob continues, ``even if you imagine that the number
on each ball is printed just after each of you presses a button, there's
still no way that the formula $F$ could be on average more than $2.$
That's because we'll make sure that each of you presses your chosen
buttons simultaneously, and as you are on opposite sides of the globe,
you can trust that no communication has been exchanged between the
boxes about which button you pressed''.

\subsubsection{The causalist's decision\label{sub:causalist}}

Alice tries to think carefully about it. She has watched this show
many times, and knows that almost everyone who takes the challenge
walks home with a million dollars. But whatever she does, the numbers
on the balls are fixed, and there's nothing she can do about that.
Bob's description of the mechanism of the boxes \emph{explains} why
the correlations most people observe can be such that $\langle F\rangle>2.8$,
but the numbers that will come out of each box cannot \emph{causally
}depend on what she does now. Alice has read about causal decision
theory, and decides to base her decision on it.

The first thing she needs to do is to calculate the \emph{causal probability
}that $\langle F\rangle>2.8$, given each sequence of button presses
at her disposal. She really has $2^{N}+1$ choices available: to press
the buttons in any of the possible $2^{N}$ combinations of red-green,
or not to press the buttons.

Alice trusts that the numbers in the marbles are determined in advance
of their choices. Or at least she trusts that the outcome of each
individual box cannot depend on the choice of experiment made in the
associated box in the other city, since there is no way a signal could
communicate that information between the boxes. Let us denote by $K_{\lambda}$
a causal hypothesis that encodes those values. In other words, $K_{\lambda}$
determines the values of all numbers on the marbles. She understands
that the boxes may be predicting what she is going to choose, but
according to causal decision theory she can't take that correlation
into account in the calculation of causal probabilities. Therefore
the causal probability for her to obtain a particular pair of values
for a pair of boxes where she chooses to press, say, the red button
(we will denote this choice by $A_{R}$) and Charlie chooses to press,
say, the green button (we will denote this by $B_{G}$) is

\begin{equation}
P_{c}(a_{r},b_{g}|A_{R},B_{G})=\sum_{\lambda}P(K_{\lambda})\, P(a_{r}|A_{R};K_{\lambda})P(b_{g}|B_{G};K_{\lambda}),\label{eq:causal_prob_marble_game}\end{equation}
following Eq.~\eqref{eq:prob_causal_A_B}. Denoting by $\langle a_{r}\rangle_{\lambda}$
the expectation value of $a_{r}$ given the variable $K_{\lambda}$,
i.e. $\langle a_{r}\rangle_{\lambda}=P(a_{r}=1|A_{R};K_{\lambda})-P(a_{r}=-1|A_{R};K_{\lambda})$,%
\footnote{Here I allow for those values to be determined probabilistically by
the $K_{\lambda}$ for generality. Of course, in the deterministic
case we simply have that \foreignlanguage{british}{$P(a_{r}=1|A_{R};K_{\lambda})\in\{0,1\}$}
and so on.%
} and similarly for the other values, we thus obtain for each of the
boxes\begin{equation}
\langle a_{r}b_{g}\rangle_{c}=\sum_{\lambda}P(K_{\lambda})\,\langle a_{r}\rangle_{\lambda}\langle b_{g}\rangle_{\lambda}.\label{eq:causal_expect_value_arbg}\end{equation}
Now Alice of course attributes the same underlying distribution of
$K_{\lambda}$ to all of the pairs of boxes. After all, as far as
she is concerned, they are all identical. Therefore, the causal expectation
value $\langle a_{r}b_{g}\rangle_{c}$ for the subset of boxes where
she chose to press the red button and Charlie chose the green button
will be just that given by Eq.~\eqref{eq:causal_expect_value_arbg}.
This will be the same, therefore, regardless of what is the subset
of boxes for which they choose the combination red-green. The same
argument tells us that the causal expectation value of $\langle a_{r}b_{r}\rangle_{c}$,
$\langle a_{g}b_{r}\rangle_{c}$ and $\langle a_{g}b_{g}\rangle_{c}$
will also be given by an equation of the form \eqref{eq:causal_expect_value_arbg},
and will be independent of Alice's and Charlie's particular choices.
And therefore the causal expectation value of $\langle F\rangle_{c}$
will be \[
\langle F\rangle_{c}=\sum_{\lambda}P(K_{\lambda})\,\left[\langle a_{r}\rangle_{\lambda}\left(\langle b_{r}\rangle_{\lambda}+\langle b_{g}\rangle_{\lambda}\right)+\langle a_{g}\rangle_{\lambda}\left(\langle b_{r}\rangle_{\lambda}-\langle b_{g}\rangle_{\lambda}\right)\right].\]
The argument of section \ref{sub:Mechanism-underlying-the} tells
Alice that each of the terms in square brackets is at most $2$, and
therefore the value of $\langle F\rangle_{c}$ can also be at most
$2$.%
\footnote{It would be possible, of course, for the actual value of $\langle F\rangle$
to be larger than $2$. But we assume for simplicity that the number
of boxes is large enough that she should not reasonably expect large
deviations from the expected value.%
} Alice therefore decides not to play the game and takes home the risk-free
thousand dollars.

The causalists could object that the example is not directly analogous
to the original Newcomb problem. I do not claim it is. But CDT should
not be valid only for the original Newcomb problem. It should be able
to be applied consistently to every decision problem, given the agents'
beliefs about the causal structure of the world. With the interpretation
about the causal structure given in the problem, this analysis leads
to the prescription exemplified above. In any case, I will present
below a modified Newcomb problem that is closer to the marble boxes
game, so as to sharpen the reader's intuition with this scenario.

There are a million closed boxes, and an open box with \$1000. Alice
can pick one and only one of the closed boxes. She can also take home
the open one if she so chooses. If the Predictor predicted Alice would
choose just one closed box instead of a closed box and the open one,
then the Predictor has also made a prediction about which one of the
million closed boxes Alice would pick and has put the money into that
box. If Alice was predicted to pick the open box as well, the million
dollars have been placed in one of the closed boxes at random. Her
evidential expected utility given that she picks only a closed box
is much larger than that given that she picks also the open box, because
she believes the Predictor is sufficiently accurate. Suppose Alice
decided to pick only box 3679, say, and found \$1 million in it. Can
she consistently believe (without a belief in retrocausality) that
the fact that box 3679 contained \$1 million was caused by her choice
to pick it? Of course not. Box 3679, she believes, already contained
\$1 million dollars before she picked it. It contained a million dollars
because the Predictor put it there. And she put it there because she
predicted Alice would pick that particular box.

Now suppose Alice is a causal decision theorist. She believes her
choice cannot cause the contents of the boxes to change. For all she
knows, there is a million dollars inside one of the closed boxes,
but she doesn't know which one. Even though her evidential probability
that it will be behind the box that she chooses is high---if she chooses
just one of the closed boxes---CDT says she can't take that correlation
into account in her decision any more than she can take into account
the correlation in the standard two-box Newcomb problem. The causal
probability that she gets a million dollars given that she picks one
of the closed boxes is given by Eq. \eqref{eq:Newcomb_P_c}, with
an unconditional average over the possible states of the boxes---i.e.,
it is 1 in a million. Therefore the causal expectation value for picking
just one of the closed boxes is just \$1, and CDT says she should
take the \$1000.

\subsubsection{The Bell game\label{sub:The-Bell-game}}

As the reader familiar with Bell's theorem already noticed, the marble
boxes scenario can be arranged with a pair of entangled quantum systems.
Bob's formula is just the expression on the left side of the Bell-CHSH
inequality \citep{Clauser1969}, which has the limit of $2$ within
local hidden variable models. The red/green buttons play the role
of the type of measurement to be performed on each particle, and the
numbers on the balls represent the two possible outcomes of these
measurements, which can be given the values of $\pm1.$ However, quantum
mechanics allows the value of $\langle F\rangle$ to be as much as
$2\sqrt{2}$ for pure entangled states of the pair of particles, with
appropriate measurement settings.

I will argue that every causalist---even those who know about Bell's
theorem---should bet against the predictions of quantum mechanics
in the Bell game%
\footnote{Here I will point out the other line of criticism mentioned in the
beginning of Section \ref{sub:Consequences-of-the}. A causalist who
has a strong belief in local causality, perhaps due to lack of knowledge
of Bell's theorem, and who is presented with a story similar to that
of the marble boxes game, would not be able to modify their bets even
in the face of a losing history, as discussed in Section \ref{sub:Evidential-and-effective}.
The evidence acquired with the losing history could change their subjective
probabilities about the situation, but those are effectively useless
as far as their gambling commitments are concerned. Given their best
theories about the causal structure of the world (presumably the causal
structure implied by the theory of relativity, with forwards causality)
the only explanation for the correlations in the marble boxes game
is in fact a common cause for their choices and the numbers in the
marbles. Since this hypothesis explains the correlations, more data
about the correlations cannot change the agents' prior beliefs about
the causal structure. They would maintain their decisions even when
faced with strong evidence that the correlations can indeed be such
that it would be advantageous to play the game, just as they would
have to maintain their decision to pick two boxes in Newcomb's problem
regardless of how much evidence they acquire about the correlation
between their choices and the contents of the closed box.

This criticism may require a particular type of causalist, perhaps
a non-existent type. A causalist that may have fallen in this category
may argue with hindsight that they have simply been cheated in this
game, and that their decision theory is not at fault. I won't try
to disagree with this conclusion. However, this discussion seems to
point at another problem for the causalist, i.e. to explain how their
causal hypotheses are modified by evidence. It would be a challenge
to the causalist's refusal to revise their causal stories in the face
of the evidence in the Newcomb scenario. That is, it would be a challenge
to explain in which sense the Newcomb scenario is different from the
quantum case that no amount of evidence (apart from \textquotedbl{}inside
information\textquotedbl{} about the workings of the Predictor's system)
can justify a modification of the causal probabilities in the Newcomb
case, whereas it can change our causal hypotheses about the quantum
case.

The causalist could here reply that their causal stories \emph{may
}be affected by evidence, and choose to emphasise that in the quantum
scenario, local causality \emph{cannot }explain the quantum correlations,
and that is the reason it was rejected as a causal hypothesis. However,
there are logically possible causal hypotheses compatible with the
quantum correlations and in which local causality is maintained in
one way or another (either through retro-causality \citep{Price1996,Pegg2008,Price2008,Berkovitz2008},
or through what Bell called {}``superdeterminism'' \citep{Bell1987}).
Given that local causality is, according to Bell and others, the causal
requirement of relativity---our best theory of causal structure---and
given the availability of alternatives that maintain it, local causality
should not be outright rejected by the causalist. Furthermore, a commonly
held position is the idea that the quantum correlations are \emph{acausal},
thus implying that the causal effects (if any) of an action like choosing
among a number of alternative experiments are still restricted to
those effects which can be caused locally. This causal hypothesis
also leads to difficulties for CDT as I will argue in more detail
in this section.

After the preparation of this manuscript it was brought to my attention
that a proposal for a Newcomb-type problem using quantum mechanics
was published in the PhD thesis of Joseph Berkovitz \citeyearpar{Berkovitz1995}.
While the underlying motivation is similar, the specific setup and
the analysis are substantially different from those of the present
work. Importantly, the arguments given by Berkovitz are aimed towards
an agent without a knowledge of Bell's theorem, as in the line of
argument mentioned above.%
}. . However, there is a subtlety: there are several alternative causal
hypotheses for the correlations given by quantum mechanics, some of
which would lead to the conclusion of section \ref{sub:causalist},
some of which wouldn't. The prescription of CDT will depend on the
causal hypothesis entertained by the agent. The problem here is that
there is no consensus about what the causal structure underlying quantum
mechanics is. Let us analyse some of the possibilities:
\begin{enumerate}
\item \textbf{Acausal correlations}. A common view is that quantum correlations
do not involve causation---they are \emph{acausal}. This view seems
to present a problem for CDT. If the correlations are acausal, then
it must be the case that the distant measurement outcome is \emph{not}
caused by the local choice of measurement (it is not caused by anything
for that matter). This view would presumably maintain that any \emph{causal}
effects are still restricted to the future light cone of an action
(otherwise there would be no reason for the unusual claim that quantum
correlations are acausal), and we should obtain for the causal probabilities
allowed in this hypothesis the same mathematical form as \eqref{eq:prob_causal_A_B}.
\item \textbf{Nonlocal causality}. The view that there is some form of nonlocal
causation involved in quantum correlations (e.g., Bohmian mechanics).
In this case, causal probabilities would coincide with the quantum
probabilities.
\item \textbf{Superdeterminism}. This is the hypothesis that local causality
is still valid, but that the independence assumption fails, i.e.,
the hidden variables are correlated with the choices of measurements
in much the same way as in the marble boxes example. Bell \citeyearpar{Bell1987}
referred to this possibility as \textquotedbl{}superdeterminism\textquotedbl{}.
This hypothesis has very low credibility in general---even though
it is the only one fully compatible with forwards relativistic causality,
as far as I am aware---since it would require conspiratorial correlations.
It is a logical possibility however, as acknowledged by Bell himself.
In this case the causal probabilities would be given by Eq.~\eqref{eq:prob_causal_A_B}.
\item \textbf{Retrocausality}. Depending on whether one also assumes local
causality, this hypothesis could be formally indistinguishable from
Hypothesis $3$, but would postulate retrocausality as an explanation
for the correlations \citep{Price1996}, and thus a failure of the
independence assumption. The causal probabilities would then be the
same as the quantum probabilities.
\end{enumerate}
Being logical possibilities, all of which with some advantages (and
disadvantages) over the others, a rational agent should ascribe some
credence, even if very small, to each of these possible causal hypotheses.
Certainly Hypothesis 3, and I believe also Hypothesis 1, both lead
to a situation where causal probabilities are given by Eq.~\eqref{eq:prob_causal_A_B}
and thus diverge from the quantum probabilities in general. In the
case of Hypothesis 3, that situation is clear, as it is analogous
to the marble boxes example.\emph{ }Either way, some nonzero credence
(call it $\epsilon$) should be assigned to some hypotheses where
the causal and quantum probabilities differ.

How should CDT deal with these different hypotheses? The obvious approach
is to understand the causal variables $K_{\lambda}$ in Eq. \eqref{eq:causal_p_OW_1}
as really representing \emph{two} variables: one that describes which
general causal hypothesis one is considering (e.g., 1, 2, 3 or 4 above)
and the second describing the actual causal variables within each
hypothesis (e.g., the hidden variables within 3). Grouping together
all causal hypotheses according to the prescribed causal probabilities
(according to whether or not they allow agreement with the quantum
probabilities), and representing the credence on causal probabilities
of the form \eqref{eq:prob_causal_A_B} by $\epsilon,$ we arrive
at

\begin{multline}
P_{c}(a_{j},b_{l}|A_{i},B_{k})=\epsilon\,\sum_{\lambda}P(K_{\lambda})\, P(a_{j}|A_{i};K_{\lambda})P(b_{l}|B_{k};K_{\lambda})\\
+(1-\epsilon)\,\sum_{\xi}P(K_{\xi})P(a_{j},b_{l}|A_{i},B_{k};K_{\xi}),\label{eq:weighted_causal_prob-1}\end{multline}
where I have used the subscript $\lambda$ to indicate the causal
variables associated with Hypothesis 3 (or any hypothesis that leads
to the same form for causal probabilities, such as perhaps Hypothesis
1 as argued above), and $\xi$ to indicate causal variables associated
with hypotheses that allow the causal probabilities to be equal to
the quantum prescription.

As shown in section \ref{sub:causalist}, the causal probabilities
prescribed by the first term will lead to a bound of $2$ on $\langle F\rangle$,
whereas the other term would be bounded by the maximal quantum-mechanical
expectation value of $2\sqrt{2}$. The weighted causal expectation
value of $\langle F\rangle$ would therefore be $\langle F\rangle_{c}\leq2\epsilon+2\sqrt{2}(1-\epsilon).$
For any nonzero value of $\epsilon,$ therefore, that expectation
value can never reach $2\sqrt{2},$ and Bob can always formulate a
game analogous to the marble boxes game simply by changing the boundary
between the region where Alice gets the million-dollar payoff and
that in which she gets nothing. With enough statistics (and this can
always be arranged in principle) the actual expectation value can
be confidently above this bound, leading to the same situation as
in the example. Therefore this argument against causal decision theory
does not depend on a strong belief in any causal hypothesis, but merely
on the acceptance that each of those are \emph{logical }possibilities
which have\emph{ some} merit (and therefore must be given some nonzero,
even if arbitrarily small, credence).

It can be argued against this conclusion that one usually assumes
that we are allowed to ignore extremely unlikely hypotheses in our
decisions. Consider, say, the hypothesis that having a cup of tea
would result in the destruction of the universe%
\footnote{This example was brought to my attention by an anonymous referee.%
}. Surely, the argument goes, we don't need to consider all logically
possible hypotheses?

My response to this criticism is that we don't consider all possible
hypotheses because we make a pre-judgement that no further hypotheses
would change our decisions, and that further considerations would
only introduce unnecessary complications in the calculations. Most
tea drinkers attribute an exceedingly small probability for the destruction
of the universe conditional on their drinking tea. But if a tea drinker
were to give any appreciable probability to this hypothesis, it would
certainly be irrational for them to have that cup of tea.

Further, in a situation like the referee's example, not only would
these kinds of unlikely hypotheses have negligible effects on the
decisions, but there would usually be equally arbitrary competing
hypotheses pulling the decision the other way: the hypothesis that
NOT having a given cup of tea will lead to the destruction of the
universe is just as (un)likely as the one that having that cup of
tea will do so, and precisely cancels the effect of the first.

There is also an important difference between the tea example and
the hypothesis of superdeterminism. Not only there are more reasons
to believe the latter---and therefore it should have a larger, even
if still small, credence---but the argument holds for any nonzero
value attributed to this credence. Besides, the causal hypotheses
underlying quantum mechanics do not affect the observable evidential
probabilities, whereas the tea hypothesis changes the expected probabilities
and utilities of the possible outcomes.

In any case, it is not necessary to hang onto the idea that one should
assign nonzero probabilities to every logically possible causal hypothesis.
All that is needed is that one assigns some nonzero credence to Hypothesis
3, or to Hypothesis 1 with the analysis above. Since our best theory
of causal structure is Einstein's theory of relativity, which according
to Bell's analysis implies local causality, and given that there are
options available to explain the quantum predictions while upholding
local causality (and either violating statistical independence or
believing in acausal correlations), it seems that we have reasons
not to completely reject those hypotheses.

An option for the causalist is to bite the bullet and maintain, with
Lewis, that the riches in the Bell game are reserved for the irrational
just as much as in the Newcomb game. This would then, interestingly,
provide a practical means to distinguish between different causal
hypotheses underlying quantum mechanics. Depending on their credences
on the various alternatives, different agents who act according to
CDT would have different thresholds for which they would accept to
play the Bell game. Their gambling commitments at different formulations
of the game would therefore be evidence of their credences on underlying
causal hypotheses, all of which being otherwise empirically indistinguishable.
In other words, varying degrees of belief in metaphysical hypotheses
would surprisingly lead to different prescriptions for the explicit
behaviour of rational agents.

\section{Communicated vs. non-communicated predictions\label{sec:Communicated-vs.-non-communicated}}

The scenario proposed in this paper brings up an interesting discussion
which I find worth mentioning here. The existence of a further knowledgeable
agent has been sometimes considered as intuition pumps for the Newcomb
problem. Schlesinger \citeyearpar{Schlesinger1974}, for example,
considered the existence of a perfectly knowledgeable well-wisher
agent who can see the contents of the closed box and give advice to
the agent in the original Newcomb problem. Clearly this well-wisher
would always advise the agent to choose both boxes. Isn't it in the
best interest of the agent to follow the advice of a well-wisher who
has more knowledge than herself about the situation?

Note, however, that whatever the contents of the box, the advice of
the well-wisher would always be the same. Therefore the well-wisher
does not convey any new information to the agent, despite the appearances.
This is the crucial point: for a Newcomb problem to exist in the first
place, the information about the relevant causal factors cannot be
available \emph{to the choosing agent}, regardless of whether it is
available to other agents. If this information was available, nothing
could prevent the agent from choosing so as to falsify the prediction;
the prediction therefore could not be accurate independently of the
agent's choice, in contradiction with the premise of the problem.
No metaphysical \textquotedbl{}free-will\textquotedbl{} is necessary
for this conclusion. All that is necessary is for an agent to be a
physical system able to carry out the deterministic algorithm: \textquotedbl{}if
I find that I was predicted to do $A_{1},$ I do $A_{2},$ and vice-versa\textquotedbl{}.

In fact, we could imagine, quite reasonably, that artificial-intelligence
(AI) programs could be constructed to take decisions in situations
where they could be caught up in a Newcomb problem. After all, there
is no mystery involved in predicting the actions of a program; all
you need is to run another copy of it with the same inputs%
\footnote{ An argument for one-boxing involving computer simulations can be
found in \citep{Aaronson2005}.%
}. Perhaps these programs could \textquotedbl{}know\textquotedbl{}
that they are in a Newcomb problem: some of the inputs could be a
prediction of the program's own output, which the programs could empirically
observe to be accurate at better than even odds as required. Suppose
a company is trying to decide whether to build their AI agents with
causal or Bayesian decision theory. Clearly the preferred algorithm
in this case would be BDT. 

There may be a way to block this conclusion: perhaps the AI could
have access to all the relevant parts of its own algorithm. It would
therefore be able to know in advance what it would have been predicted
to choose. This would not mean, however, that the AI would be able
to perform better in Newcomb problems. On the contrary, it would mean
that it would not be able to find itself in Newcomb problems in the
first place. This is the essence of the \textquotedbl{}tickle defence\textquotedbl{}
of BDT. It was first proposed by Horgan \citeyearpar{Horgan1981}
in reference to the \textquotedbl{}smoking gene\textquotedbl{} problem
reproduced in Section \ref{sec:Causal-Decision-Theory}. His argument
was that in this case the flaw is in \textquotedbl{}the assumption
that the agent needs to \emph{act} before he has the relevant information
to determine the likelihood of getting lung cancer. He does not, because
his own desirabilities (and past behaviour) give him the bad news
already.\textquotedbl{} With that information, Horgan argues, the
agent can screen off the conditional probabilities such that the probabilities
of having or not the gene (and therefore of developing lung cancer)
are independent of his choice. There are some more sophisticated variations
on this type of argument \citep{Eells1982,Price1986,Price1991}, but
the general idea is that a careful account of a rational agent's capabilities
would block the need for CDT, as Newcomb-type problems would not be
generally feasible, and BDT would therefore always give the same prescriptions
as CDT. This could be due to the fact that a rational agent acts according
to her beliefs and desires---there are no further relevant causal
factors to affect her decisions. To the extent that she knows her
beliefs and desires, she knows all that the Predictor could know in
order to predict her choices, and therefore she knows the prediction.
As a consequence, she cannot find herself in a Newcomb problem, since
she would not believe in the defining conditional probabilities.

But this only solves the problem by dismissing it, and makes CDT simply
irrelevant, as there would be no problem in which the causal and evidential
probabilities are different. Horgan admits, however, that the problem
could be (implausibly) reformulated so that, say, the genetic factor
in question induces in smokers a tendency to choose to continue smoking
when faced with this problem. Horgan concedes that this avoids the
tickle defence, but believes that, in this case, it would actually
be rational to stop smoking, even though that decision does not \emph{cause}
the desirable outcome. 

Within a hidden-variables interpretation, the quantum scenario makes
it possible to instantiate an actual Newcomb-type problem by making
it impossible even in principle for an agent to know the hidden variables.
If this were not the case, then where Bell violations occur, an agent
could use their knowledge of the hidden variables to transmit faster-than-light
signals \citep{Cavalcanti2007t}, which could lead to the existence
of inconsistent causal loops. The fact that the probabilities in a
usual Newcomb scenario seem to be apparently independent of whether
or not the agent could have knowledge of the causal factors speaks
against the feasibility of those scenarios. Any actual instantiation
of a Newcomb-type problem would probably look much more like the Bell
scenario than the medical Newcomb problems. They should make it clear
that those causal factors are in fact hidden as far as the agent is
concerned.

\section{Summary and conclusion}

The main argument of this paper can be summarised as follows: (i)
CDT needs some account of which events are within and outside the
causal influence of an action; (ii) with this distinction in place,
the causal probabilities are formally identical to a LHV model in
the context of quantum mechanics; (iii) the causal decision theorist
should assign some nonzero credence to the logically possible causal
hypotheses listed in \ref{sub:The-Bell-game}; (iv) the causal probabilities
for some of those hypotheses are given by a LHV model and are thus
distinct from the quantum probabilities. A game can be constructed
to exploit that discrepancy, following Bell's theorem; (v) since any
observation is by construction compatible with all of the causal hypotheses,
repeated observation of the predicted quantum correlations cannot
change the initial credences and CDT will always prescribe the losing
strategy.

Perhaps this debate may also inform discussions in foundations of
quantum mechanics. The fact that the debate in decision theory centres
around the \textquotedbl{}statistical independence\textquotedbl{}
assumption may indicate that this assumption, often taken for granted,
needs more attention in the quantum debate. One way of relaxing that
assumption is in terms of a kind of superdeterministic theory in which
both the experimental outcomes and choices share a common cause. Another
possibility is that these correlations are arranged through retrocausality.
Some authors have considered this possibility as a serious alternative
to the interpretation of quantum mechanics \citep{Price1996,Wharton2007,Berkovitz2008,Pegg2008,Price2008},
but it hasn't been given as much attention as it seems to deserve.

Decision theory has been used as the basis of a foundational program
started by Deutsch \citeyearpar{Deutsch1999} and further developed
by Wallace \citeyearpar{Wallace2006,Wallace2007} in the context of
the Everett (Many-Worlds) interpretation of quantum mechanics. However,
that parallel could also prove useful in attempts to understand quantum
mechanics as a theory about information \citep{Caves2002b,Fuchs2003}.
One of the goals of this program is to pursue information-theoretic
principles that lead one to the abstract formalism of quantum mechanics.
The discussion in this paper seems to indicate that it might be interesting
for that program to consider the perspective of an agent not only
as the holder of information, but as the source of decisions about
observations to be performed on the world. These decisions, as far
as the agent is concerned, cannot be considered to be correlated with
any of \emph{their} information (which experiments an agent will perform
is not encoded in their quantum state assignment), but yet their observations
are such that the world looks \emph{as if} the outcomes of those observations
were so correlated with their choices, if only they consider the general
validity of local causality\emph{. }Regardless of commitments about
the actual existence or otherwise of hidden variables, it would be
interesting to know whether these kinds of considerations can restrict
the space of possible theories in an interesting way.

As topics for further research, it would be interesting to attempt
to find simpler decision scenarios displaying an incompatibility between
CDT and BDT within a quantum set up. A possible approach would be
to use the correlations of a Greenberger-Horne-Zeilinger tri-partite
entangled state, which allow for non-statistical demonstrations of
the incompatibility between local realism and the predictions of quantum
theory, or an adapted form of the Bell-Kochen-Specker theorem.

Finally, as discussed in Section \ref{sub:The-Bell-game}, if causal
decision theory is the correct theory of rational decisions, then
this analogy would provide a surprising practical consequence, in
terms of the prescribed behaviour of rational agents, for competing
causal interpretations of quantum mechanics. In other words, it would
provide an observable, practical distinction between alternative metaphysical
hypotheses.

\paragraph*{Acknowledgements}

I would like to thank Joseph Berkovitz, Rachael Briggs, Julia Bursten,
Phil Dowe, Huw Price and Howard Wiseman for useful discussions and
feedback. This work was partly supported by an Australian Research
Council Postdoctoral Research Fellowship.

\bibliographystyle{bjps}
\bibliography{/Users/eric/Documents/Work/bib/Eric}

\end{document}